\documentclass[english,aps,pra,reprint,superscriptaddress]{revtex4-1}
\usepackage[T1]{fontenc}
\usepackage[latin9]{inputenc}
\usepackage{verbatim}
\usepackage{float}
\usepackage{textcomp}
\usepackage{amsmath}
\usepackage{amssymb}
\usepackage{graphicx}
\usepackage{color}

\makeatletter

\newcommand{\lyxmathsym}[1]{\ifmmode\begingroup\def\b@ld{bold}
  \text{\ifx\math@version\b@ld\bfseries\fi#1}\endgroup\else#1\fi}

\usepackage{braket}

\makeatother

\usepackage{babel}
\begin{document}
\title{Floquet band engineering and topological phase transitions in 1T' transition metal dichalcogenides}
\author{Xiangru Kong}
\email{kongx@ornl.gov}
\affiliation{Center for Nanophase Materials Sciences, Oak Ridge National Laboratory,
Oak Ridge, Tennessee 37831, USA}

\author{Wei Luo}
\affiliation{Center for Nanophase Materials Sciences, Oak Ridge National Laboratory,
Oak Ridge, Tennessee 37831, USA}

\author{Linyang Li}
\affiliation{School of Science, Hebei University of Technology, Tianjin 300401,
China}

\author{Mina Yoon}
\affiliation{Center for Nanophase Materials Sciences, Oak Ridge National Laboratory,
Oak Ridge, Tennessee 37831, USA}

\author{Tom Berlijn}
\affiliation{Center for Nanophase Materials Sciences, Oak Ridge National Laboratory,
Oak Ridge, Tennessee 37831, USA}

\author{Liangbo Liang}
\email{liangl1@ornl.gov}
\affiliation{Center for Nanophase Materials Sciences, Oak Ridge National Laboratory,
Oak Ridge, Tennessee 37831, USA}

\begin{abstract}

Using \textit{ab initio} tight-binding approaches, we investigate Floquet band engineering of the 1T' phase of transition metal dichalcogenides (MX$_2$, M = W, Mo and X = Te, Se, S) monolayers under the irradiation with circularly polarized light. Our first principles calculations demonstrate that light can induce important transitions in the topological phases of this emerging materials family. For example, upon irradiation, Te-based MX$_2$ undergoes a phase transition from quantum spin Hall (QSH) semimetal to time-reversal symmetry broken QSH insulator with a nontrivial band gap of up to 92.5 meV. On the other hand, Se- and S-based MX$_2$ undergoes the topological phase transition from the QSH effect to the quantum anomalous Hall (QAH) effect and into trivial phases with increasing light intensity. From a general perspective, our work brings further insight into non-equilibrium topological systems.\newline

\end{abstract}
\maketitle

\section{Introduction}
Typically, controlling topological phase transitions requires the
change in the structural properties of materials~\cite{bernevig2006quantum,konig2007quantum}.
Alternatively, among the many developed experimental techniques for
probing and controlling quantum systems, laser and microwave driving
fields can be used as a tool that provides time-periodic fields
to induce non-equilibrium topological phenomena in solid states, cold
atoms, and photonics systems~\cite{oka2019floquet,rudner2020band,basov2017towards,PhysRevResearch.2.043408}. Theoretically, Floquet engineering with Floquet-Bloch theory~\cite{PhysRevLett.110.200403,bukov2015universal},
which is used for analyzing periodically driven quantum systems, proves
itself as a powerful tool for studying the Floquet topological phases~\cite{lindner2011floquet,PhysRevLett.120.156406,aeschlimann2021survival}.
Basically, there are two key paradigms for inducing topological phase transitions
in quantum systems~\cite{rudner2020band}: resonant and off-resonant driving. In the resonant
driving scheme, with the frequency (energy of fields) larger than the
band gap and smaller than the bandwidth, the coupling between the
valence and conduction bands will result in expected band inversions;
In the off-resonant scheme, the driving frequency is required to be larger
than the single-particle bandwidth, and the driving fields
cannot resonantly couple states in the valence and conduction bands
for any value of momentum in the Brillouin zone; thus the band inversion
is mainly dependent on the field intensity.

Early on, graphene under the irradiation of light was proposed for realizing the quantum anomalous
Hall (QAH) effect by inducing a band opening at the Dirac point~\cite{PhysRevB.79.081406,PhysRevB.89.121401,PhysRevLett.113.266801,sentef2015theory}.
Then a nearly quantized anomalous Hall conductance was
observed in graphene experimentally~\cite{mciver2020light}.
This success triggered intense research interests in Floquet engineering in solid states, as it is still rare to realize the QAH effect in real materials.
Floquet engineering of topological phase transitions has been proposed in trivial systems such as HgTe/CdTe quantum wells~\cite{lindner2011floquet}, transition metal dichalcogenides (TMDs)~\cite{claassen2016all,Sie2014ws2,de2016monitoring}, and strained black phosphorus~\cite{PhysRevLett.120.237403}. Floquet engineering can also be applied in topological nontrivial materials such as the surface of topological insulator Bi$_2$Se$_3$~\cite{wang2013observation,mahmood2016selective}, the Dirac semimetal Na$_3$Bi~\cite{hubener2017creating}, and the antiferromagnetic FeSe monolayer with the quantum spin Hall (QSH) effect~\cite{PhysRevLett.120.156406}.

Recently, two-dimensional (2D) TMDs with chemical formula MX$_2$ (M = W, Mo and X = Te, Se, S) have
attracted much attention. The family of monolayer TMDs can crystallize into
three common  polymorphic structures: 1H (hexagonal), 1T (tetragonal)
and 1T' (distorted tetragonal). The 1T' structural phase
was predicted to host the QSH effect~\cite{qian2014quantum}, and topological properties have been experimentally observed in both 1T'-WTe$_2$ ~\cite{tang2017quantum,fei2017edge,wu2018observation} and 1T'-WSe$_2$~\cite{chen2018large,ugeda2018observation,pedramrazi2019manipulating}. Remarkably, superconductivity was also reported in 1T'-WTe$_2$ by applying
a gate voltage to change the carrier density~\cite{sajadi2018gate,fatemi2018electrically} and unconventional superconducting behaviors were found in 1T'-MoS$_2$ ~\cite{peng2019high,PhysRevB.98.184513}.
The meet of topology and superconductivity raises hope for the realization
of topological superconductors~\cite{PhysRevLett.125.107001,PhysRevLett.125.097001}.

In our work, we investigate Floquet engineering in 1T' MX$_2$ monolayers by using the $\textit{ab initio}$ Wannier-function based tight binding method~\cite{PhysRevLett.120.156406,kong2020magnetic}. The results show that there are semimetal to insulator phase transitions in Te-based MX$_2$ monolayers under light irradiation. The QSH metal phase predicted by the $\textit{ab initio}$ method can be driven into the large gap time-reversal symmetry broken QSH insulator phase by light, which will be important in the study of topological properties in Te-based MX$_2$ monolayers.
Moreover, the light-induced QAH effect can be observed in Se- and S-based MX$_2$ monolayers, further demonstrating the power of Floquet engineering in controlling and manipulating the topological properties of 1T' TMDs.

\section{Computational Methods}
$\textbf{\textit{ab initio method}}$.
First-principles calculations were employed in the Vienna \textit{ab initio }simulation
package (VASP) within the framework of density functional theory (DFT) by the projected augmented wave
(PAW) method~\cite{PhysRev.136.B864,PhysRevB.54.11169,PhysRevB.59.1758}. The electron exchange-correlation functional is chosen as the generalized
gradient approximation (GGA) of Perdew, Burke and Ernzerhof (PBE)~\cite{PhysRevLett.77.3865}. The atomic positions were relaxed by the conjugate gradient scheme
until the maximum force was less than 1 meV/$\text{\AA}$ on each atom with the lattice constants adapted from Ref.~\cite{qian2014quantum}. A large vacuum region of more than 16 $\text{\AA}$ in the $z$ direction was used to avoid
spurious interactions with the neighboring cells. The total energy
was converged to 10$^{-8}$ eV. The energy cutoff of the plane waves
was chosen as 400 eV. The k-mesh in the Brillouin Zone was sampled
using $\Gamma$-centered $9\times9\times1$ k-grid. The screened Heyd-Scuseria-Ernzerhof Hybrid functional method (HSE06)~\cite{heyd2003hybrid,paier2006screened} was also used to demonstrate the light-induced Floquet bands in WTe$_2$ systems, for which the band gap values are under debate by previous experimental and theoretical studies.

\textbf{\textit{Wannier-function based tight binding Hamiltonian.}} An effective tight-binding Hamiltonian constructed from the maximally localized Wannier function (MLWF) was used to illustrate the Floquet band structures~\cite{mostofi2014updated,kong2020magnetic} with Floquet-Bloch theory. The $d$ orbitals from transition metal (W, Mo) and the $p$ orbitals from the chalcogen atoms (Te, Se, S) were projected as the Wannier basis. The topological properties of the Floquet-Bloch systems were determined using the iterative Green's function method for surface states~\cite{wu2018wanniertools,sancho1985highly} and the Wilson loop or Wannier charge centers (WCCs) for analyzing the topological invariants~\cite{wu2018wanniertools,PhysRevB.83.235401,PhysRevB.84.075119,PhysRevB.95.075146}.

\textbf{\textit{Floquet-Bloch theory.}} We consider an external time-dependent circularly polarized
light irradiation along the out-of-plane direction of 1T' MX$_2$ as shown in Fig. $\text{\ref{fig:cartoon}}$(a).
The time-periodic light can be described as $\mathbf{A}(t)=A(\eta\sin(\omega t),\cos(\omega t),0)$,
where $\eta=\pm1$ indicates the chirality of the circularly polarized
light, $\omega$ is the frequency, $A$ is the light amplitude and
$T$ is the driving period with $\omega=2\pi/T$. Based on Floquet-Bloch
theory~\cite{PhysRevLett.110.200403,PhysRevLett.120.156406}, the
periodically driven $d+1$ dimensional quantum lattices can be described
as an effective static Hamiltonian
\begin{align}
H_{F}(\mathbf{k},\omega) & =\sum_{mn}\sum_{\alpha\beta}\alpha\hbar\omega\delta_{mn}\delta_{\alpha\beta}c_{m\alpha}^{\dagger}(\mathbf{k})c_{n\beta}(\mathbf{k})\nonumber \\
 & +\sum_{mn}\sum_{\alpha\beta}\sum_{j}\tau_{j}^{mn}\cdot\frac{1}{T}\int_{0}^{T}dte^{i[\frac{e}{\hbar}\mathbf{A}(t)\cdot\mathbf{d}_{j}^{mn}-q\omega t]}\\
 & \times e^{i\mathbf{k}\cdot\mathbf{R}_{j}}c_{m\alpha}^{\dagger}(\mathbf{k})c_{n\beta}(\mathbf{k})+h.c.,\nonumber 
\end{align}
where $\tau_{j}^{mn}$ is the hopping term between the $m$th Wannier
orbital in $0$th cell and $n$th Wannier orbital in $j$th cell with
the corresponding position vector $\mathbf{d}_{j}^{mn}$, $\mathbf{R}_{j}$
is the lattice vector, and $(\alpha,\beta)$ is Floquet index (integers)
ranging from $-\infty$ to $\infty$ with $q=\alpha-\beta$. Here, by testing our calculations, we choose $q=\pm1$ to get the converged results without considering more Floquet bands. Moreover,
we focus on the off-resonant driving paradigm with $\hbar\omega=13$
eV which is larger than any bandwidths of the projected bands in 1T' TMDs,
so the light intensity is the only parameter to tune the Floquet bands. It saves us from worrying
about the ambiguity and complexity of band inversions in the on-resonant driving paradigm where the Floquet bands can be coupled to each other~\cite{zhao2020folding}. We have also examined a lower photon energy of $\hbar\omega=7$
eV, the main conclusions of our results are still valid as shown in Fig. S11 for WSe$_2$ and Fig. S12 for WTe$_2$ in the Supplemental Material~\cite{Supplemental_Material}.

\section{Results and Discussion}

\begin{figure}[ht!]
\centering{}\includegraphics[width=1.0\columnwidth,clip]{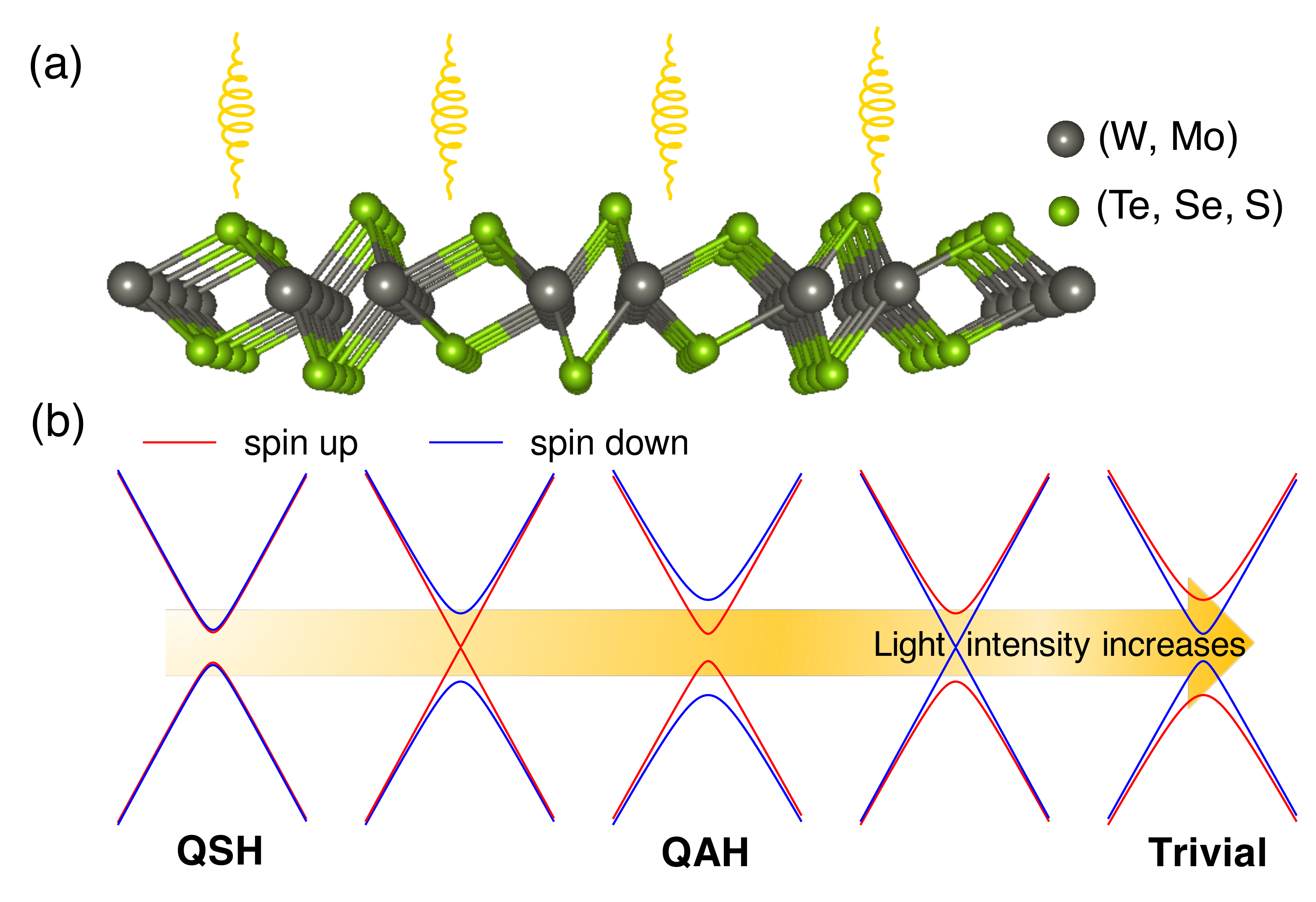}\caption{\label{fig:cartoon} (a) Side view of 1T' MX$_2$ (M = W, Mo and
X = Te, Se, S) monolayer under an external time-dependent circularly
polarized light irradiation. (b) Schematic band diagrams show
the light-induced topological phase transition from the (time-reversal symmetry broken)
QSH effect to the QAH effect, then to trivial phases.}
\end{figure}

\begin{figure}[ht!]
\centering{}\includegraphics[width=1.0\columnwidth,clip]{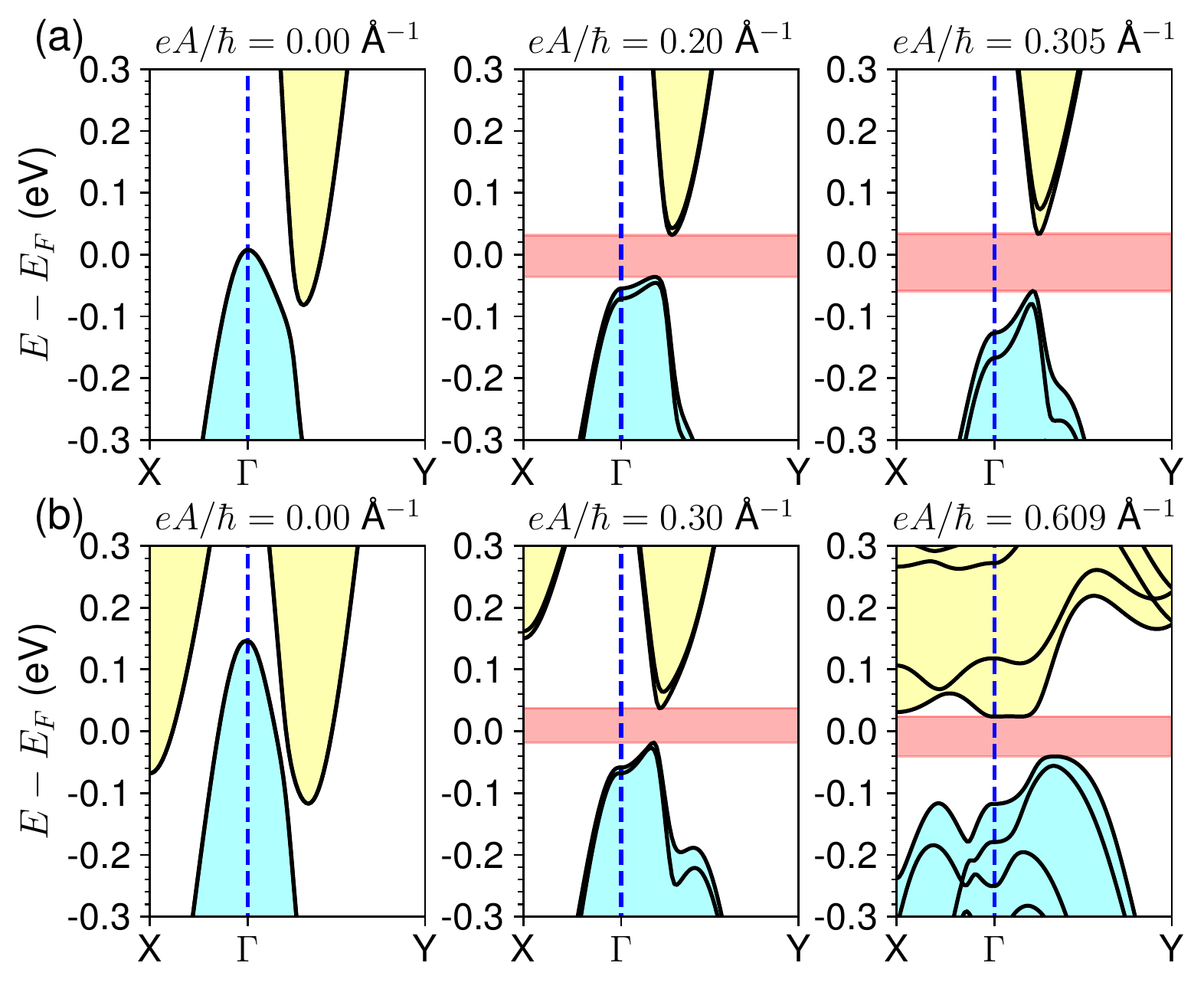}\caption{\label{fig:TMTe2}Light-induced Floquet bands in (a) 1T' WTe$_2$
with $eA/\hbar=$0.00, 0.20, and 0.305 $\text{\AA}^{-1}$
and (b) 1T' MoTe$_2$ with $eA/\hbar=$0.00, 0.30, and 0.609 $\text{\AA}^{-1}$. The red regions are plotted as the eye guidance for the full band gap.}
\end{figure}

First, we focus on the Floquet engineering in Te-based TMDs, 1T' WTe$_2$ and MoTe$_2$ (see Fig.~\ref{fig:cartoon}(a)), since the Te-based TMDs are the most energetically stable and also the most experimentally accessible among the six materials studied in this work~\cite{duerloo2014structural}. The conducting behavior of Te-based TMDs (metallic or insulating), as probed via different measurement techniques, is still under debate~\cite{zheng2016quantum,keum2015bandgap,tang2017quantum,song2018observation,hu2021realistic}.  
However, even if 1T' Te-based TMDs are insulating, enlarging the band gaps is still important for observing the topological properties.
Therefore it is still worth exploring how to enlarge the band gaps in 1T' Te-based TMDs. For example, a tendency towards gap opening is reported in 1T'-MoTe$_2$ which is in proximity to a topological insulator~\cite{zhang2021tendency}, and tuning the semimetal phase to topological insulator phase in monolayer 1T'-WTe$_2$ is also studied via using strain engineering technique~\cite{PhysRevLett.125.046801}. In these works, the exchange-correlation functional of PBE was applied to elaborate how the band gaps can be enlarged, demonstrating that the PBE functional, also used in our work, is effective to describe the electronic properties of TMDs. Here, the PBE calculations identify both 1T' WTe$_2$ and MoTe$_2$ as QSH metals (QSHM)~\cite{qian2014quantum} (see the bands without light irradiation in the left panels of Fig. $\text{\ref{fig:TMTe2}}$(a) and $\text{\ref{fig:TMTe2}}$(b)), where the slight overlap of conduction bands and valence bands indicates metallicity in the presence of the QSH effect. We also characterize their topological properties: the evolution of WCCs in half the Brillouin Zone gives the $Z_2=1$ (Fig. S1(a) and S2(a)); the gapless helical edge states, which are protected by the time-reversal symmetry, can be observed but are merged with the metallic bulk states (Fig. S1(a) and Fig. S2(a)). Therefore, it is desirable to  
eliminate the metallic bulk properties and have a finite band gap to take advantage of the topological edge states. For example, strain engineering was applied in WTe$_2$ for this purpose~\cite{PhysRevLett.125.046801}.

\begin{figure}[ht!]
\centering{}\includegraphics[width=1.0\columnwidth,clip]{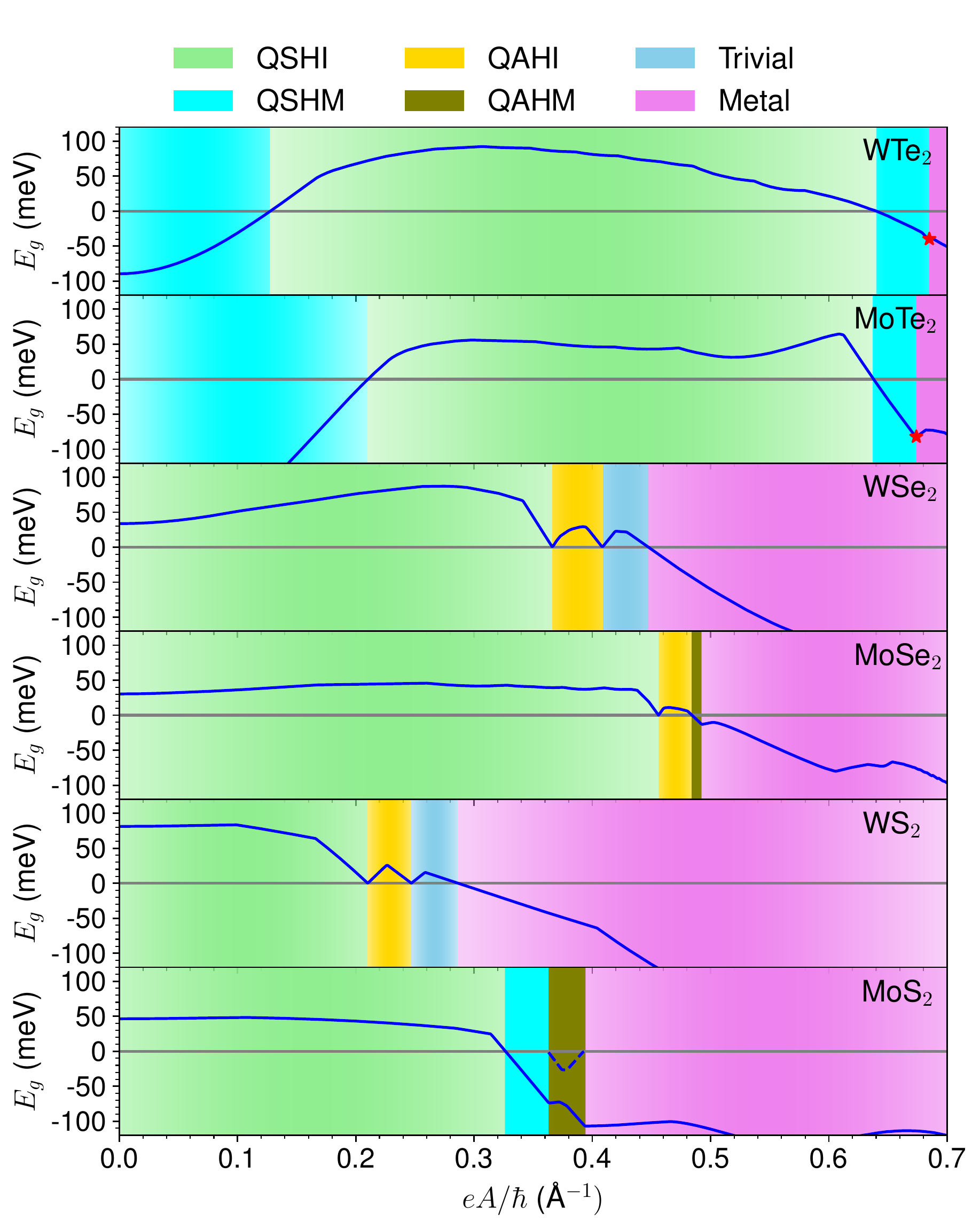}\caption{\label{fig:gaps} The evolution of Floquet band gaps ($E_g$) in the 1T' MX$_2$ as the light intensity increases. $E_g$ is defined as the difference between the minimum of conduction bands and maximum of valence bands, which are around the $\Gamma$ point along the high symmetry line $X-\Gamma-Y$. Note that the Floquet band gaps (blue dashed line) of MoS$_2$ are also defined at the $\Gamma$ point between $eA/\hbar=$0.326 $\text{\AA}^{-1}$ and $eA/\hbar=$0.394 $\text{\AA}^{-1}$ to show the band inversions, due to insulator-metal phase transitions before the topological phase transition. Different colored regions indicate different light-induced phases: QSHI(M) indicates the QSH effect with insulator (metal) properties; QAHI(M) indicates the QAH effect with insulator (metal) properties; trivial indicates the trivial insulator phase; the metal phase indicates the region with complex Floquet bands, which may contain band inversions. The red stars indicate the first band inversions around $X$ points in Te-based TMDs. The critical values are shown in Table S1\cite{Supplemental_Material}}
\end{figure}

Floquet engineering can be an alternative way to drive the QSHM phase into the full gap QSH insulator (QSHI) phase.
For example, upon light irradiation, the overlap of conduction and valence bands of Te-based TMDs decreases with increasing the light intensity (Fig. $\text{\ref{fig:gaps}}$). A further increase in light intensity drives them into full gap QSHIs.
The critical light intensity above which they become QSHI is 0.127 $\text{\AA}^{-1}$ for WTe$_2$ and 0.210 $\text{\AA}^{-1}$ for MoTe$_2$.
The light-induced band gaps continue to increase with light intensity and then decrease back to the QSHM phase (Fig. $\text{\ref{fig:gaps}}$).
For WTe$_2$, the full gap can be observed in Fig. $\text{\ref{fig:TMTe2}}$(a) with
$eA/\hbar=$0.20 $\text{\AA}^{-1}$, and the largest band gap can reach
to 92.5 meV when $eA/\hbar=$0.305 $\text{\AA}^{-1}$.
From Fig. S1(b) when $eA/\hbar=$0.305 $\text{\AA}^{-1}$,
the topological properties can still be observed from the evolution of WCCs
and the helical edge states but with small gaps around the edge Dirac points.
As discussed in Ref.~\cite{PhysRevLett.107.066602,PhysRevLett.123.096401}, the QSH effect under the light irradiation is in fact the time-reversal symmetry broken QSH effect.
The circularly polarized light irradiation breaks the time-reversal symmetry, but the system still preserves the QSH effect when the light intensity is small enough according to adiabatic
evolution theorem~\cite{RevModPhys.82.3045}.
The small gaps in the helical edge states are caused by the breaking of both time-reversal symmetry and inversion symmetry, as also observed in graphene under Zeeman field~\cite{PhysRevLett.107.066602}.
For the MoTe$_2$ system, similar phenomena can also be observed, as shown
in Fig. $\ref{fig:TMTe2}$(b) and Fig. $\text{\ref{fig:gaps}}$.
The largest light-induced nontrivial band gap is 64.5 meV 
at $eA/\hbar=$0.609 $\text{\AA}^{-1}$ (Fig. $\text{\ref{fig:TMTe2}}$(b)).
The topological properties when $eA/\hbar$=0.609 $\text{\AA}^{-1}$ are
demonstrated in Fig. S2(b). Considering that hybrid functional (HSE06) calculations had predicted Te-based TMDs to be full gap QSHIs\cite{keum2015bandgap,zheng2016quantum,tang2017quantum}, we also performed the HSE06 calculations as shown in Figs. S13 and S14. Although Te-based TMDs already have finite gaps without light irradiation at the hybrid functional level, the light can still manipulate and enlarge the nontrivial gaps of Te-based TMDs with increasing light intensity, similar to the results found by the PBE calculations discussed above. To summarize, regardless of the initial conducting state of Te-based TMDs (metallic or insulating), both PBE and HSE06 methods predict that light irradiation is a promising technique to enlarge the band gaps of TMDs, and therefore can facilitate the experimental observation of topological behaviors of Te-based TMDs\cite{PhysRevLett.125.046801,zhang2021tendency}, since larger gaps are easier for clear experimental detection.

While light can induce full band gaps in Te-based TMDs, it can also induce topological phase transitions in Se- and S-based TMDs as illustrated in Fig. $\ref{fig:cartoon}$(b): the transition from the time-reversal symmetry broken QSH effect phase to the trivial
insulator phase requires an intermediate QAH effect phase~\cite{PhysRevLett.107.066602}.
The time-reversal symmetry broken QSH effect can still be characterized by nonzero spin Chern
numbers $C_{\pm}=\pm1$, 
which are Chern number of spin up/down subspace of the occupied bands~\cite{li2013spin,PhysRevB.80.125327}.
As the light intensity increases, it should preserve the time-reversal
broken QSH effect before any band inversion happens. 
However, when light intensity increases to a critical value, band inversion
occurs in the spin-up subspace of the occupied bands due to the increasing of
the bands splitting. This leads to the spin-up Chern number
becoming zero, and induces the QAH effect characterized by the spin-down Chern number. If then another band inversion occurs in the spin-down subspace of occupied bands at or around the same $k$-point, topological nontrivial phase breaks down.

\begin{figure*}
\centering{}\includegraphics[width=1.0\linewidth,clip]{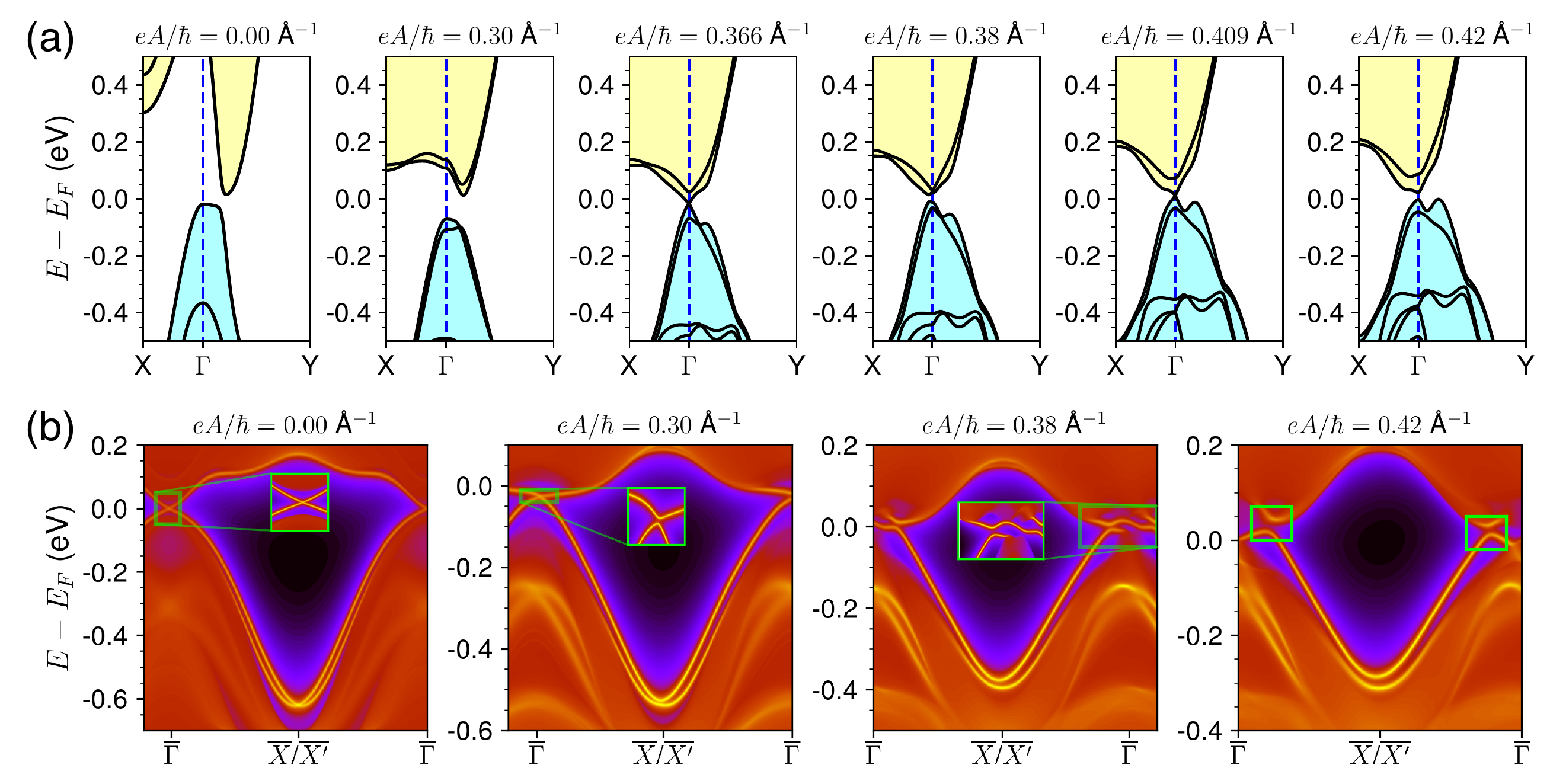}\caption{\label{fig:WSe2}(a) Light-induced Floquet bands in 1T' WSe$_2$
with $eA/\hbar=$0.00, 0.30, 0.366, 0.38, 0.409 and 0.42 $\text{\AA}^{-1}$.
(b) The edge states along the direction [10] with $eA/\hbar=$0.00,
0.30, 0.38 and 0.42 $\text{\AA}^{-1}$; the insets demonstrate the magnification of the
gapless or gapped edge states. Here, $\eta=-1$ as the left circularly
polarized light is used for the irradiation. Fermi levels are
shifted for each specific value of $eA/\hbar$.}
\end{figure*}

The schematic diagram in Fig.~\ref{fig:cartoon}(b) summarizes the essence of the topological phase transitions observed in Se- and S-based TMDs. Fig.~$\ref{fig:gaps}$ identifies the transitions in topological phases as a function of light intensity and the evolution in Floquet band gaps. The 1T' WSe$_2$ has been identified as a large gap QSHI in experiments~\cite{chen2018large,ugeda2018observation,pedramrazi2019manipulating}.
The Floquet bands of WSe$_2$ with $eA/\hbar=$0 $\text{\AA}^{-1}$ reproduces the DFT bands (see Fig.~\ref{fig:WSe2}(a)), implying that the time-reversal symmetry protected QSH effect is preserved. The gapless helical edge states are located at the $\overline{X}$ point along the [10] and [01] directions
(see Fig.~\ref{fig:WSe2}(b) and S4), and we confirm $Z_2=1$ by evaluating the evolution of the WCCs (Fig. S3).
With increasing light intensity, as shown in Fig.~$\ref{fig:gaps}$, the Floquet band gaps increase slightly. 
The band splittings near the Fermi level can be observed as shown in Fig.~\ref{fig:WSe2}(a) when  $eA/\hbar=0.3$ $\text{\AA}^{-1}$. The evolution of WCCs at $eA/\hbar=0.3$ $\text{\AA}^{-1}$ in Fig. S3 indicates the time-reversal symmetry broken QSH effect, and the helical edge states with small gaps can be observed in Fig. $\text{\ref{fig:WSe2}}$(b) with the insets.
When $eA/\hbar=0.366$ $\text{\AA}^{-1}$, the first band gap closing is observed
as shown in Fig. $\text{\ref{fig:WSe2}}$(a), and then the gap opening
occurs. This apparent indication of band inversion implies that there
is a topological phase transition.
As the light intensity continues to increase to $eA/\hbar=0.38$ $\text{\AA}^{-1}$, 
the evolution of WCCs along the whole Brillouin Zone in Fig. S3 demonstrates that the Chern number $C=-1$, indicating that the system becomes a QAH insulator (QAHI).
We also confirm that the system exhibits a chiral edge state across $\overline{\Gamma}$ point (see Fig. $\text{\ref{fig:WSe2}}$(b) and S4).
If we increase the light intensity further, another band inversion
can be observed at $eA/\hbar=0.409$ $\text{\AA}^{-1}$ in Fig. $\text{\ref{fig:WSe2}}$(a).
This indicates another topological phase transition. However, it should
be a nontrivial to trivial phase transition, as this is the second band inversion occurring at the same $k$-point. This is confirmed by both the evolution of WCCs in Fig. S3 and the edge states
in Fig. $\text{\ref{fig:WSe2}}$(b) and Fig. S4.

The light-induced topological phase transition in WSe$_2$ is consistent with Fig. $\text{\ref{fig:cartoon}}$(b). If we look further at the
evolution of band gaps in WSe$_2$ in Fig.~$\ref{fig:gaps}$, the full band gap can be increased
from 33.7 meV to 87.2 meV when the light intensity $eA/\hbar$ increases
from 0 to 0.274 $\text{\AA}^{-1}$. Then, as discussed above and shown
in Table S1, when $0.366<eA/\hbar<0.409$ $\text{\AA}^{-1}$, the Floquet-Bloch
system behaves as a QAHI with the largest gap of 29.3 meV. Finally, 
the system transitions into a trivial insulator phase, and then
into a trivial metal phase with the increasing light intensity.
Meanwhile, the WS$_2$ system under the light irradiation exhibits
the same feature of phase transitions (Fig. $\text{\ref{fig:cartoon}}$(b)),
and the critical values of topological phase transitions are 0.210
and 0.247 $\text{\AA}^{-1}$ (Table S1) as shown in Fig.~$\ref{fig:gaps}$. The evolution of Floquet
bands and WCCs in WS$_2$ are demonstrated in Figs. S5 and
S6. 

However, there is a slight difference in MoSe$_2$ and MoS$_2$
systems. In MoSe$_2$, the time-reversal
broken QSHI phase transitions into the QAHI phase at $eA/\hbar=0.456$ $\text{\AA}^{-1}$, but before the second band inversion occurs, the overlap of conduction and valence bands
makes it become a QAH metal (QAHM) phase after $eA/\hbar=0.484$ $\text{\AA}^{-1}$ as shown in the narrow region of Fig.~\ref{fig:gaps}.
Then at $eA/\hbar=0.492$ $\text{\AA}^{-1}$, there is the second band
inversion which transitions the QAHM phase into the trivial metal phase directly. The
evolution of Floquet bands and WCCs in MoSe$_2$ are demonstrated
in Figs. S7 and S8, and we can clearly observe the overlap of conduction
and valence bands in Fig. S7 which results in a non-full gap phase,
i.e., the trivial metal phase. For MoS$_2$ under the light irradiation,
the insulator to metal phase transition happens before the
topological phase transition as shown in Fig.~\ref{fig:gaps} and Table S1. After
$eA/\hbar=0.326$ $\text{\AA}^{-1}$, the QSHI phase transitions into the
QSHM phase, and we observe the metal phase in Fig. S9 when $eA/\hbar=0.328$ $\text{\AA}^{-1}$.
Also as shown in Fig. S9, the overlap of conduction and valence bands
not only makes the full band gap disappear, but also renders the local band gap difficult to be observed. However, by tracking the band gap only at the $\Gamma$ point (blue dashed line in Fig.~$\ref{fig:gaps}$) and the evolution of
WCCs in Fig. S10, we can still locate the critical values of topological
phase transitions as shown in Table S1: 0.363 and 0.394 $\text{\AA}^{-1}$. Therefore, the differences in the phase diagrams between Se- ans S- based TMDs (see Fig.~\ref{fig:gaps}) are due to the occurring order of the light-induced topological phase transition and insulator-metal (or metal-insulator) phase transition. 

Notably, though the topological phase transitions
are not observed at the relatively low light intensity in Te-based TMDs, it can still
be possible if we further increase the light intensity. As shown in
Fig.~\ref{fig:gaps}, after Te-based TMDs transition into the QSHM phase from the QSHI phase, increasing light intensity will further enhance the overlap of conduction
and valence bands, until there is a band inversion happening. We track
the evolution of light-induced band gaps, and observe that for
Te-based TMDs the first band inversion occurs around the $X$ point as the red stars in Fig.~\ref{fig:gaps},
not the $\Gamma$ point as in Se- and S-based TMDs. Moreover, after
the first band inversion, Te-based TMDs under the light irradiation
are in the QAHM phase, but as the light-induced Floquet bands become more
complex, the second band inversion happens around other $k$ points
other than the $X$ point. This makes the situation complicated, and the topological phase transition mechanism
in Fig. $\text{\ref{fig:cartoon}}$(b) does not apply.

\begin{figure}
\begin{centering}
\includegraphics[width=1.0\columnwidth,clip]{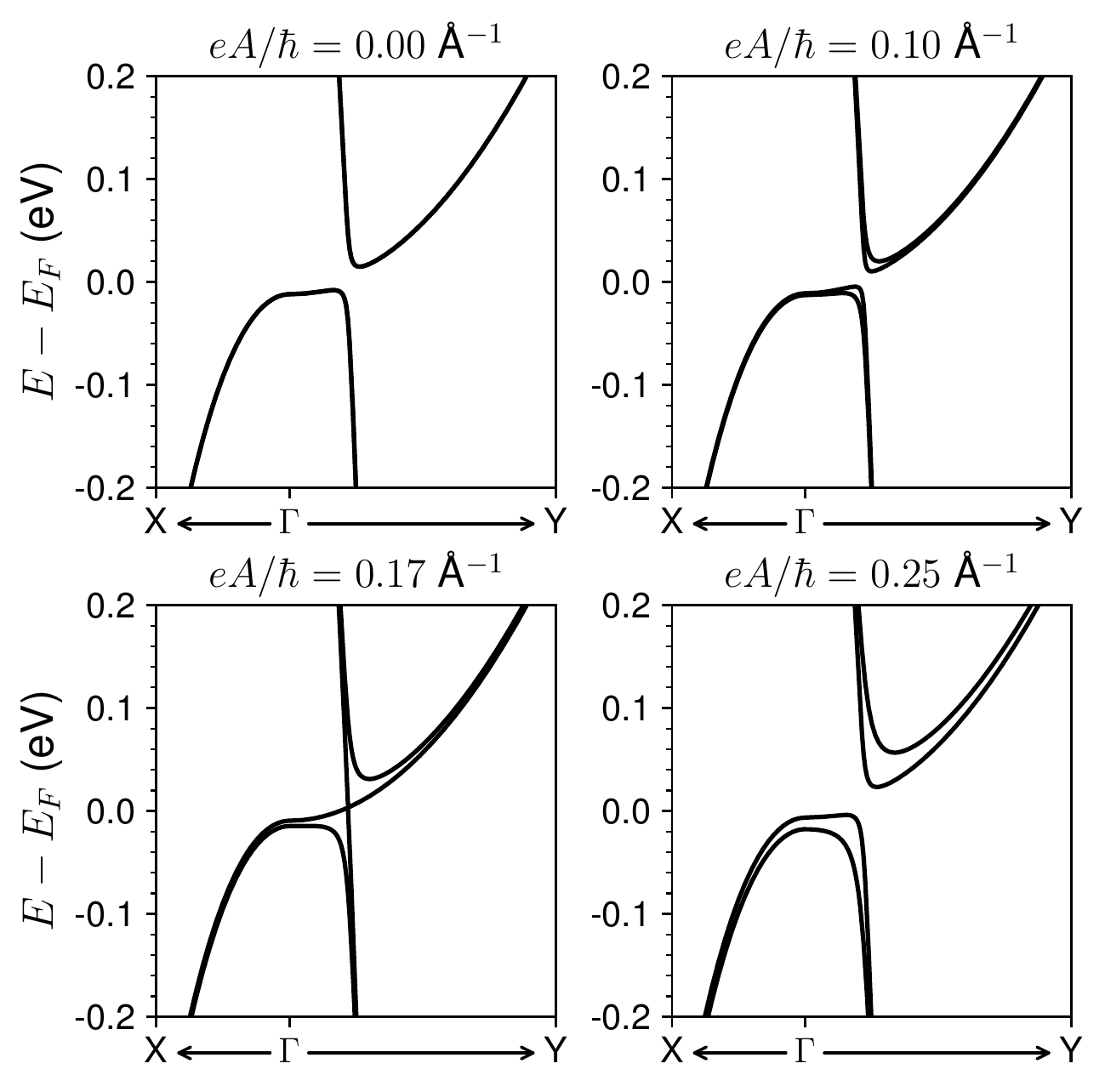}\caption{\label{fig:kp} The evolution of Floquet bands within a four band
$k\cdot p$ model for 1T' WSe$_2$. High frequency ($\hbar\omega=13$ eV) left
circularly polarized light ($\eta=-1$) is used for the irradiation.}
\par\end{centering}
\end{figure}

Finally, to demonstrate how the light can induce band inversions, we use a four band
$k\cdot p$ model designed for 1T' TMDs ~\cite{qian2014quantum,das2020tuneable} to further study the Floquet band engineering. By solving the corresponding $H_{eff}$ (see more details in the Supplemental Material~\cite{Supplemental_Material}), we can observe the band inversion with the
increasing light intensity as shown in Fig. $\text{\ref{fig:kp}}$.
The first order of high frequency expansion, $[H_{-1},H_{1}]/\hbar\omega$,
introduces an effective spin orbital coupling (SOC) term. This
effective SOC term results in the band splittings as shown in
Fig. $\text{\ref{fig:kp}}$. With the increasing light intensity,
the gap opening and closing can be captured. Therefore, it is clear that the band inversion can be elucidated
by the effective Hamiltonian, which is consistent with the results
of the \textit{ab initio} Wannier based tight-binding method discussed above.

\section{Conclusion}

In conclusion, our systematical study based on $\textit{ab initio}$ calculations reveals noteworthy light-induced phase transitions in 1T' MX$_2$ (M = W, Mo and X = Te, Se, and S) under the irradiation of an external time-dependent circularly polarized light. The light-induced metal to insulator phase transition in Te-based TMDs (i.e., 1T' WTe$_2$ and MoTe$_2$) demonstrates that the full and large band gap QSH effect can be obtained by Floquet engineering.
Moreover, the light-induced topological phase transitions in Se- and S-based TMDs, which undergo band inversions twice around the $\Gamma$ point, shows that the time-reversal symmetry broken QSH effect phase must transitions into the QAH effect phase first before the trivial phases. Therefore, 1T' MX$_2$ can be possible platforms for realizing quantized anomalous Hall conductance via the light irradiation. This study demonstrates the advantage of Floquet engineering in two-dimensional systems for inducing and controlling topological phase transitions, and is expected to inspire further research interests in the field of non-equilibrium topological systems.

\begin{acknowledgments}

We thank Mark Dean, Yue Cao, Rob Moore and Hu Miao for valuable discussions. This research was conducted at the Center for Nanophase Materials
Sciences, which is a DOE Office of Science User Facility. We used
resources of the Compute and Data Environment for Science (CADES)
at the Oak Ridge National Laboratory, which is supported by the Office
of Science of the U.S. Department of Energy under Contract No. DE-AC05-
00OR22725. Linyang Li acknowledges financial support from the National Natural Science Foundation of China (Grant No. 12004097), the Natural Science Foundation of Hebei Province (Grant No. A2020202031), and the Foundation for the Introduced Overseas Scholars of Hebei Province (Grant No. C20200313).\newline

\indent Notice: This manuscript has been authored by UT-Battelle, LLC under Contract No. DE-AC05-00OR22725 with the U.S. Department of Energy.  The United States Government retains and the publisher, by accepting the article for publication, acknowledges that the United States Government retains a non-exclusive, paid-up, irrevocable, world-wide license to publish or reproduce the published form of this manuscript, or allow others to do so, for United States Government purposes.  The Department of Energy will provide public access to these results of federally sponsored research in accordance with the DOE Public Access Plan (\textcolor{blue}{http://energy.gov/downloads/doe-public-access-plan}).

\end{acknowledgments}





\bibliography{references}

\end{document}